\documentclass[conference]{IEEEtran}
\IEEEoverridecommandlockouts
\usepackage{cite}
\usepackage{amsmath,amssymb,amsfonts}
\usepackage{algorithmic}
\usepackage{graphicx}
\usepackage{textcomp}
\usepackage{xcolor}
\usepackage[ruled,vlined]{algorithm2e}
\usepackage{enumitem}
\makeatletter
\newcommand{\linebreakand}{%
  \end{@IEEEauthorhalign}
  \hfill\mbox{}\par
  \mbox{}\hfill\begin{@IEEEauthorhalign}
}
\makeatother
\begin{document}

\title{An Analytics Framework for Heuristic Inference Attacks against Industrial Control Systems}

\author{
\IEEEauthorblockN{Taejun Choi}
\IEEEauthorblockA{\textit{The University of Queensland}\\
Brisbane, Australia \\
taejun.choi@uq.edu.au}
\and
\IEEEauthorblockN{Guangdong Bai}
\IEEEauthorblockA{\textit{The University of Queensland}\\
Brisbane, Australia \\
g.bai@uq.edu.au}
\and
\IEEEauthorblockN{Ryan K L Ko}
\IEEEauthorblockA{\textit{The University of Queensland}\\
Brisbane, Australia \\
ryan.ko@uq.edu.au}
\linebreakand
\IEEEauthorblockN{Naipeng Dong}
\IEEEauthorblockA{\textit{The University of Queensland}\\
Brisbane, Australia \\
n.dong@uq.edu.au}
\and
\IEEEauthorblockN{Wenlu Zhang}
\IEEEauthorblockA{\textit{The University of Queensland}\\
Brisbane, Australia \\
wenlu.zhang1@uq.net.au}
\and
\IEEEauthorblockN{Shunyao Wang}
\IEEEauthorblockA{\textit{The University of Queensland}\\
Brisbane, Australia \\
shunyao.wang@uq.net.au}
}

\maketitle

\begin{abstract}

Industrial control systems (ICS) of critical infrastructure are increasingly connected to the Internet for remote site management at scale. However, cyber attacks against ICS -- especially at the communication channels between human-machine interface (HMIs) and programmable logic controllers (PLCs) -- are increasing at a rate which outstrips the rate of mitigation. 

In this paper, we introduce a vendor-agnostic analytics framework which allows security researchers to analyse attacks against ICS systems, even if the researchers have zero control automation domain knowledge or are faced with a myriad of heterogenous ICS systems. 
Unlike existing works that require expertise in domain knowledge and specialised tool usage, our analytics framework does not require prior knowledge about ICS communication protocols, PLCs, and expertise of any network penetration testing tool. Using  `digital twin' scenarios comprising industry-representative HMIs, PLCs and firewalls in our test lab, our framework's steps were demonstrated to successfully implement a stealthy deception attack based on false data injection attacks (FDIA). Furthermore, our framework also demonstrated the relative ease of attack dataset collection, and the ability to leverage well-known penetration testing tools. 

We also introduce the concept of `heuristic inference attacks', a new family of attack types on ICS which is agnostic to PLC and HMI brands/models commonly deployed in ICS. Our experiments were also validated on a separate ICS dataset collected from a cyber-physical scenario of  water utilities. Finally, we utilized time complexity theory to estimate the difficulty for the attacker to conduct the proposed packet analyses, and recommended countermeasures based on our findings.
 
\end{abstract}


\section{Introduction} \label{section1}
Modern power grids, water treatment/distribution systems, and automated manufacturing systems are some of the main critical infrastructure sectors adopting industrial control systems (ICS) for automation and remote plant management capabilities. In the recent decade, ICS are increasingly connected to the Internet to achieve scale and management efficiency, as compared to traditional ICS management which do not leverage Internet connectivity. As a result, several ICS equipment which were previously not designed for the threat landscape of the Internet are now increasingly exposed to risks of cyber attacks. 

These risks affect business continuity requirements and may have safety implications. Since they were not designed or deployed with Internet connectivity in the first place, legacy and deployed ICS are mostly not ready for this increasing trend in ICS security threats. Worse, to secure the ICS equipment (e.g. programmable logic controllers (PLCs)), one would usually require additional expertise about the targets and specialized/bespoke systems set up before one can analyse or test them. Such a resource-consuming process has largely hindered effective security research, especially for security researchers who lack prior knowledge about the ICS system under analysis.  

To allow security testing to catch up with the rate of threats to ICS, having effective frameworks for stakeholders (e.g. asset/equipment owners) and security researchers to analyse and test for security vulnerabilities without the need for specific automation or configuration domain knowledge of heterogenous equipment would be desirable. 

An emerging research area in ICS security is threat detection. In the case of detection approaches, there have been recurring obstacles (re)producing or obtaining attack data. For example, research conducted by Hadziosmanović et al. \cite{10.1145/2664243.2664277} and  Lan et al. \cite{9045827} are examples of researchers producing attack data to evaluate their suggested approaches. Nevertheless, 
%ambiguities in the description of their attack techniques 
reproducing data based on their attack techniques may lead to additional research activities and obstacles for newcomers in this area.
% TJ Camera 'date' -> 'data'
In other cases where attack datasets are shared among multiple institutions or researchers, 
%the lack of specific system architecture knowledge and attack scenario for producing the datasets
the lack of knowledge on the specific system architecture and attack scenarios in which the datasets were produced  
significantly limit research activities and research translation potential.

To mitigate the above gaps and provide a solution for acquiring attack datasets, we propose an analytics framework. While the framework is extensible, in this paper, we will focus on the false data injection attack (FDIA) \cite{7377479,5122-5127} on ICS network packets to illustrate its capabilities. 
%Using well-known network security assessment tools, researchers might be able to replicate and conduct the proposed framework without any prior knowledge on specific PLC models and ICS under analysis. 
The proposed framework could be easily applied and reproduced using well-known publicly-available network security assessment tools, without any prior knowledge on the specific PLC models and ICS under analysis.

We coin our testing attack type as a \textit{`heuristic inference attack'} -- since attackers will be able to infer PLC signals commands through a range of heuristic-driven analysis approaches over our framework. To carry out the heuristic inference attack, the network request and response packets from the Human-Machine Interfaces (HMI) and PLCs are sorted by their periodical sending order, and then classified according to their packet lengths. 
%As we observed that similar-length data packets are used for the communication between HMI and PLCs, we have chosen packet length as a criterion for classification. 
The reason we have chosen packet length as a criterion for classification is out of our observation that similar-length data packets are used for the communication between HMI and PLCs.  
In addition, we discovered that some particular query packets from HMI have been sent periodically and the response packets from PLCs against a particular query packet have predictably similar lengths. Our proposed framework was based on these discovered traits and features of PLC and HMI communication.

The main contributions of this research are:
\begin{itemize}
\item We propose an analytics framework to provide solutions to ease attack dataset collection. Our proposed method is a stealthy deception attack based on FDIA and it could be conducted using well-known penetration testing tools. 
\item We introduce `heuristic inference attacks', a new family of attack types on ICS which is agnostic to PLC and HMI brands/models commonly deployed in ICS.
\item We estimate the difficulty for the attacker to conduct the proposed packet analysis. The time complexity theory is used to support this difficulty assessment.
\item Some recommendations of countermeasures are proposed based on the findings of 
%this research result.
our research. 
\end{itemize}

In Section \ref{section2}, background materials are presented and we provide the design of our framework in Section \ref{section3}. We discuss case studies including an explanation of the used simulation system in Section \ref{section4}. Result analyses and countermeasures are discussed in Section \ref{section5} and Section \ref{section6} shows related work. We then conclude and propose future work in Section \ref{section7}.

\section{Background} \label{section2}

\begin{figure}[htbp]
\centerline{\includegraphics[width=8cm, height=8cm]{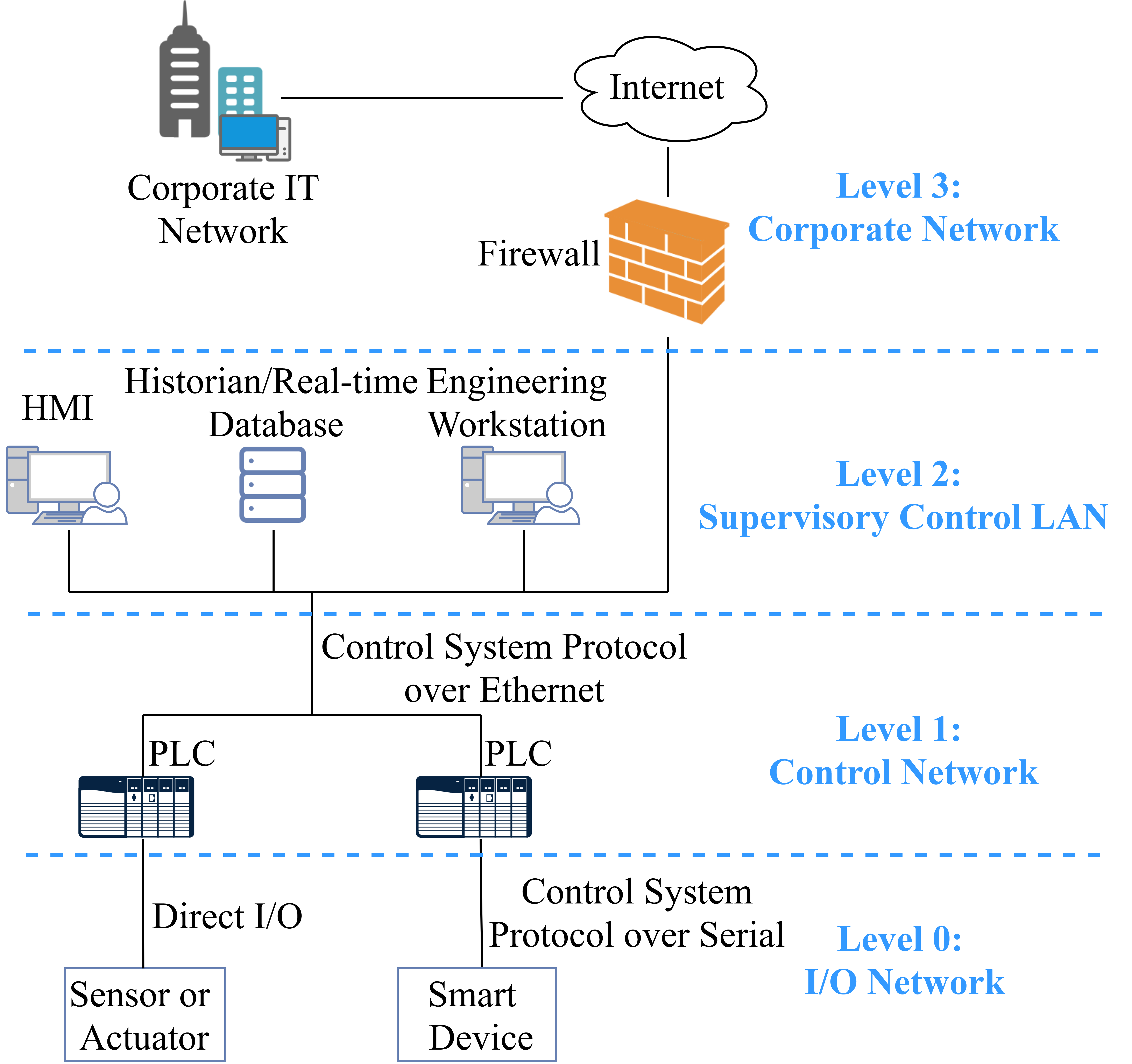}}
\caption{ICS Reference Model from ANSI/ISA-99\cite{6846667}}
\label{fig1}
\end{figure}
% Shunyao

% need more explanation about the Fig. 1 by level.
From a network perspective, ICS are usually distributed into four levels: Corporate Network (Level $3$), Supervisory Control LAN (Level $2$), Control Network (Level $1$), and I/O Network (Level $0$). All levels cooperate together to control industrial processes \cite{6846667}. Figure 1 shows a prevalent ICS architectures stemming from the ANSI/ISA-99 reference model \cite{6846667}. Level $3$ of the model is part of the traditional `Information Technology (IT)' which reports process data from lower levels to the corporate network and shares corporate data with the entire system, while Level $2$ to $0$ are classified as the `Operational Technology (OT)' which monitors and controls the actual industrial components. In particular, Level $1$ consists of PLCs which are programmed to control production processes and send obtained data from the I/O network (Level $0$) to the managing components at the supervisory control LAN (Level $2$) as described in \cite{kalle2019clik}. The HMI, Historian database, and Engineering Workstation at Level $2$ are components of the control center for monitoring and managing the OT levels.
% Wenlu

Among the four network levels, the network communication between Level $2$ and $1$ is the focus of our analysis and the exploitation target/scope of this research, since the communication channel between Level $2$ and $1$ is a necessary pathway of network attack for adversaries. For Level $3$, since the approach of attacking the OT network from the IT network has been developed by Klick et al.\cite{klick2015internet},
it is out of scope of this research. In the case of attacking Level $0$, attackers are unable to access the devices in the I/O Network or Control System without physical access to them. More importantly, the gained data from I/O devices, such as sensors or actuators, are delivered to PLCs, and then passed to the control center. Therefore, a malicious accessor of the OT network is able to gather the level $0$ data from the reported packets. Thus, attacking Level $0$ has been implicitly included.
Contrary to previous studies which focused on communications within the same network level, our approach is more reflective of real-world risks as we focus on the communication channel between HMI in Level $2$ and PLCs in Level $1$. 
%describe how attackers get into an OT network from Level 3 IT network, getting into Level 2 is not a topic to be tackled in this research and will be another research topic of ICS security. Hence, this research focuses on the communication channel between HMI in Level 2 and PLCs in Level 1.

Unlike existing works that require expertise in domain knowledge and tool usage, our framework does not require prior knowledge about ICS communication protocols, PLCs, and expertise of any network penetration testing tool. 
%\textcolor{red}{@TJ: check the following two sentences}
For example, the analysis proposed by Bernieri et al. \cite{bernieri2017monitoring} requires knowledge about the ICS systems' architecture under analysis and their relevant communication protocols. 
Similarly, in-house tools have been used for \cite{kleinmann2017stealthy}. Notably, some tools utilized for \cite{yrlmaz2019cyber} are not available any more. All of these increase the complexity in producing attack data on ICS network for security research. 
From a security assessment perspective, choosing the right tools for a particular attack scenario is the first hurdle, and the lack of experience using the chosen tools is the second obstacle.
To reduce these obstacles, our framework extensively uses off-the-shelf tools that are publicly available and well documented. In addition, our proposed framework is comprehensive enough to enable analysts without much expertise to conduct them. It further addresses the following two main challenges in communication protocol analyses: 
\begin{itemize}
\item Figuring out meaningful network packets or protocols. This is challenging because the captured packets from the OT network are mixed for different usage. 
\item Choosing the right data field of the chosen protocol packets to analyse, as the potential researchers might not have prior knowledge or expertise of ICS. 
\end{itemize}
%
%an undergraduate or master student with guidance from their supervisors. 
A detailed approach to handle these two challenges are introduced in Section \ref{section4}. We estimate that our framework will be straightforward for average-skilled researchers to use (see Section~\ref{section5}).

%Although our framework does not require prior knowledge of ICS, some challenges exist in committing the proposed steps. 

%In the case of protocols, there are two challenges to be solved. 
%In Section \ref{section4}, we introduce how we handled the described challenges. 

\section{Design of Analytics Framework} \label{section3}

%To address some huddles for researchers mentioned in \ref{section1} and \ref{section2}, we propose an attack framework against ICS in this section.
In this section, we introduce our proposed analytics framework, and start by discussing the threat model. A threat modelling of ICS was conducted to decide the research areas related to the appropriate tools used for conducting the experiment. Following that, we discuss how we applied our heuristic inference method -- which has been well developed in web security domain \cite{10.1145/1809100.1809107} -- into ICS protocols to address the aforementioned research gaps. 

\subsection{Threat Model}
Our threat model is summarised from our study of real-world security incidents. Therefore, in this section, we first present the real-world threats against ICS, before presenting our threat model. 
\subsubsection{Real Threats}
One of the most prominent ICS breaches by a disgruntled ex-contractor who was able to retain insider access to ICS systems occurred in 2000 at the Maroochy Water Services at Queensland’s Sunshine Coast in Australia. According to \cite{10.1007/978-0-387-75462-8_6}, the communication channel using radio links were lost, and pumps and alert alarms were not working properly. In 2010, Stuxnet, a malicious worm targeting Iranian nuclear enrichment plants was found to propagate through USB devices or internal networks. As Langner mentioned in \cite{5772960}, Stuxnet targeted particular Siemens PLC models to hijack their control. 
In 2015, remote administration tools (RATs) and levelled operating system or ICS client software were used to manipulate victims' systems. A 2016 alert \cite{alert2016cyber} reported that remotely accessed adversaries used these tools to cause huge blackout incidents across Ukraine. FireEye reported in 2017 that the TRITON malware caused the failed state of Safety Instrumented System (SIS) controllers. As stated by \cite{di2018triton}, attackers started to attack the system from the IT level using traditional ways then moved into the OT level.

\subsubsection{Summarized Attack Model}
Based on the above real-world attack incidents, we classify three attack models. 
\paragraph{Intrusion into OT network} To get into the OT network, adversaries have used a variety of tactics, such as malicious insiders, social engineering, and well-known attack methods in the IT network, as described in \cite{10.1007/978-0-387-75462-8_6, 5772960, di2018triton}. In addition, there were studies describing how researchers could take over Level 3 to get into the Level 2 network as described in \cite{klick2015internet}.
Therefore, since we can assume that the researchers or attackers are able to access the OT network in this research because of the existence of related researches and attack cases, attack scenarios for Level 3 are out of scope for this paper.
%check later
\paragraph{Man-in-the-Middle (MitM) Attacks} MitM attacks are commonly found in communication channels \cite{stricot2016taxonomy}. As \cite{5772960} describes that the attacker intercept interactions between PLCs and I/O modules, we assume that researchers are able to capture network packets between the HMI server and PLCs by MitM attacks. To commit this attack, attackers may deceive a victim HMI server and PLCs using an attack machine. They send some messages from the attack machine to victim PLCs to make them sending packets for HMI to the attack machine. The same deception technique could be used on a victim HMI.
%TJ One of the well-known operating systems for this penetration is Kali-Linux. As a further assumption, we presume that researchers are familiar with the operating system to use it for the MitM attack.
\paragraph{Monitoring HMI}\label{Monitoring_HMI} Like the real-world scenarios above, we consider that attackers will be able to monitor HMI using existing remote administration tools or ICS client software. Controlling an HMI server enables an in-depth analysis of the victim ICS by an attacker. As \cite{RATs_ICS} mentioned, these tools are widely used for monitoring the HMI in industrial networks. The Ukraine blackout incident was an example of such attack. 

%and the Ukraine Blackout incident used such tools was the case committed by malicious breakers.    

\subsection{Analytics Framework}
After describing our attack models, we now propose analytics steps for conducting FDIA based on the modeled threats. These steps are device-agnostic and applicable to various types of PLCs manufactured by different vendors. We have conducted the FDIA similar to the one presented in \cite{kleinmann2017stealthy}, but with more sophisticated steps for achieving attack flexibility.

\begin{figure}[t]
\centerline{\includegraphics[width=6cm, height=8cm]{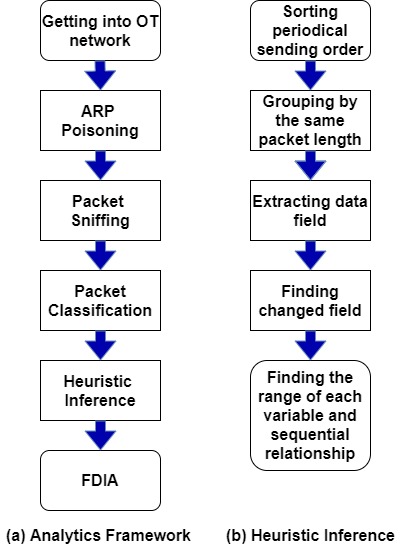}}
\caption{Analytics Framework and Heuristic Inference Steps}
\label{fig2}
\end{figure}
% Wenlu

\subsubsection{Simple Analytics Framework}
Figure 2(a) illustrates the steps of our proposed analytics framework -- from `Get into OT network' to `FDIA'.
We briefly describe each step below. Details such as the specific tools used in each step can be found in Section \ref{section4}.

%\begin{itemize}
\begin{description}
\item [Framework Step 1]: \textit{`Getting into OT network'} is the first step of the analytics framework.
% TJ Camera 'omitted' -> 'omit'
We omit the detailed method of how to conduct this step, and refer the reader to the related research discussed in Section \ref{section2} for how to infiltrate into the internal network of ICS.

\item [Framework Step 2]: \textit{`ARP poisoning'} is a relatively straightforward way to execute the MitM attack as described in \cite{kang2015investigating}. Here, researchers are able to commit the MitM attack between an HMI server and PLCs as a preliminary stage for sniffing network packets.

\item [Framework Step 3]: Gathering data by the \textit{`Packet sniffing'} step is the essential step for analysing the targeted ICS. By using any publicly available network protocol analyser, network packets between the HMI and PLCs could be recorded into the researchers' data storage.

\item [Framework Step 4]: The \textit{`Packet classification'} step determines the bundles of data packets for conducting detailed investigations/analyses at the next step. The source IP address and protocol types are the classification criteria considered first.

\item [Framework Step 5]: The \textit{`Heuristic inference'} step aids researchers to derive the targeted data field of network packets for modifying the field. Section \ref{HI} details our approach. 

\item [Framework Step 6]: Through the \textit{`FDIA'} step, researchers can produce fake network data attacking targeted HMI servers or PLCs. Manipulating the data field chosen to inject false data is in this step. Methods to modify the data field have been presented by existing studies including \cite{7377479,bernieri2017monitoring,kleinmann2017stealthy}, and \cite{kang2015investigating}.
\end{description}					
%\end{itemize}

\subsubsection{Heuristic Inference Steps}\label{HI}
In this section, we detail our heuristic inference approach. The commitment of heuristic inference to figure out the data fields for manipulation is one of the essential steps of FDIA, even though this step is usually not easy for researchers who have no prior knowledge about ICS and its protocols. We divided the inference approach into the following steps.

\begin{description}
\item [Inference Step 1]: \textit{`Sorting periodical sending order'}.
As \cite{9045827} and \cite{7377479} have pointed out, %the periodic packet exchange between PLCs and their server, 
an HMI server sequentially and periodically sends request packets with a particular length for each purpose to a PLC. When replying to these requests, PLCs would send packets with a specific length to the server according to the request type. For instance, if the sequence of request packet length is `100, 120, 80, 80, 50, 100, 120, 80, 80, 50, 100, ...', then the PLC replies to the server with packet lengths in order of `70, 90, 110, 110, 60, 70, 90, 110, 110, 60, 70, ...'. The captured data packets could be sorted using this consistent feature of ICS internal communication.

% TJ Camera 
\item [Inference Step 2]: \textit{`Grouping by the same packet length'}.
Next, we take the network packet length as a characteristic for grouping network packets between PLC and HMI, similar to \cite{9045827}. Our stragetery is based on the observation of same-length network packets periodically sent inside the OT network, as the example in the previous step shows. %\cite{9045827} shows packet length as the packet basic characteristic in their research.

\item [Inference Step 3]: \textit{`Extracting data field'}.
As mentioned by Farooq et al. \cite{9032992}, the Modbus/TCP protocol could be used for communication between the corporate and control network, the HMI and the PLC use the TCP protocol to encapsulate ICS protocol data (e.g., Modbus, Common Industrial Protocol (CIP)). Therefore, the data field, after filtering out the meaningless header field from network packets, is required.

\item [Inference Step 4]: \textit{`Finding changed field'}.
We observed that the major part of the data field of network packets have the same packet length. We hypothesized that this fact is related to the periodic transmission of same-length request/response network packets between HMI and PLCs. Also, it is assumed that the variable field might be related request/response fields querying or reporting the status of the I/O modules controlled by the PLC. Based on this hypothesis, we conclude that the request/report value from HMI or PLC can be stealthily manipulated by figuring out of the variable field.

\item [Inference Step 5]: \textit{`Finding the range of each variable and sequential relationship'}.
After figuring out the variable data fields, conducting an investigation to find the range of each variable and sequential order of changes on them is required. Since ICS is generally composed of multiple devices, the system has a sequence order at the activation of devices and the status change. If an adversary attacks an ICS system by violating the order and the range for a data field, it will be identified by a monitoring agent of the system. Therefore, this step is added to our framework to achieve unnoticeable/undetectable FDIA.
   
\end{description}

\section{Experiments and Case Studies} \label{section4}
Our proposed analytics framework was tested at the UQ Industry 4.0 Energy TestLab, our lab with a comprehensive range of `digital twin' systems mimicking real-world scenarios. In this research, we focus on a scenario of a wind-farm powering automation (represented by a barrel conveyor belt) in a manufacturing plant (see Figures 3 and 4). The digital twin scenario is configured by automation engineers and physically hosted in our lab. In this section, we first introduce our `digital twin' system, and then describe our experiment applying our framework to the digital twin. 
\subsection{Digital Twin}

The test digital twin system comprises of an HMI, two PLCs, two network switches, and a firewall -- representative components in most ICS scenarios. For the engineering workstation, a Dell OptiPlex 7070 desktop with InTouch software from Schneider Electric installed has been used for the implementation of a HMI digital twin. Figure 4 shows screen captures of the HMI. An Allen-Bradley CompactLogix 5370-L3 simulates a conveyor belt and a Schneider Electric Modicon M580 is used as a wind farm digital twin providing electricity to the conveyor belt. To simulate the firewall, pfSense, an open-source firewall, was ported onto a commercial system board. 

\begin{figure}[t]
\centerline{\includegraphics[width=8cm, height=8cm]{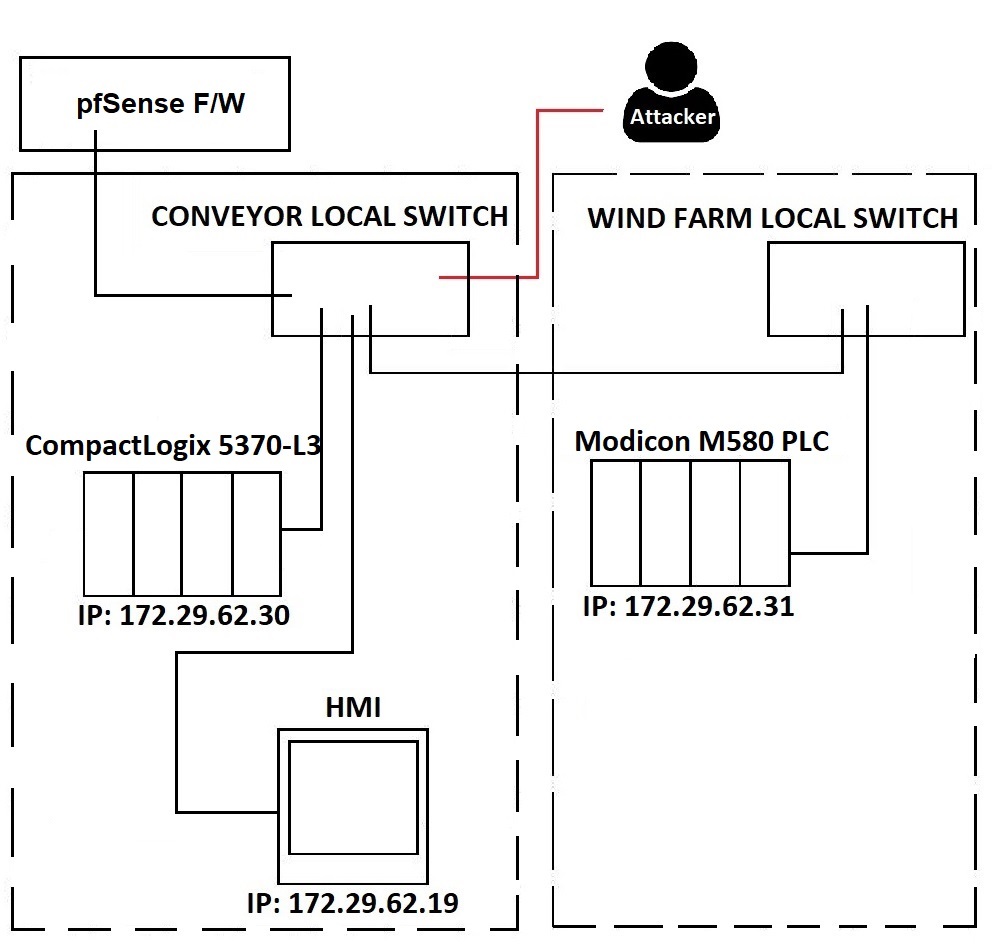}}
\caption{Simulation System Structure}
\label{fig3}
\end{figure}
% Wenlu

\begin{figure}[htbp]
\centerline{\includegraphics[width=8cm, height=8cm]{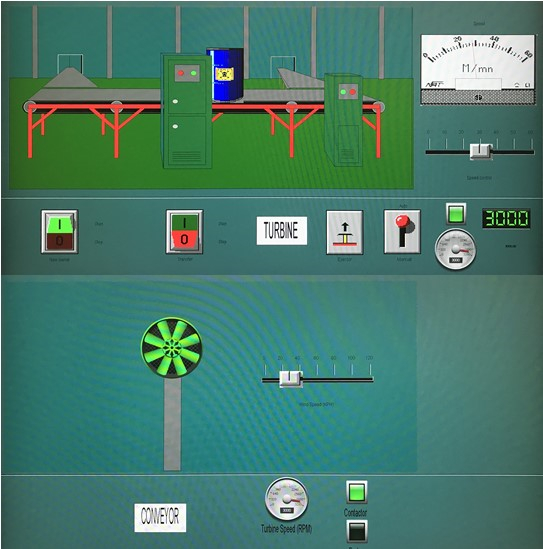}}
\caption{Human Machine Interface}
\label{fig4}
\end{figure}
% Shunyao
\begin{figure}[htbp]
\centerline{\includegraphics[width=9cm, height=5cm]{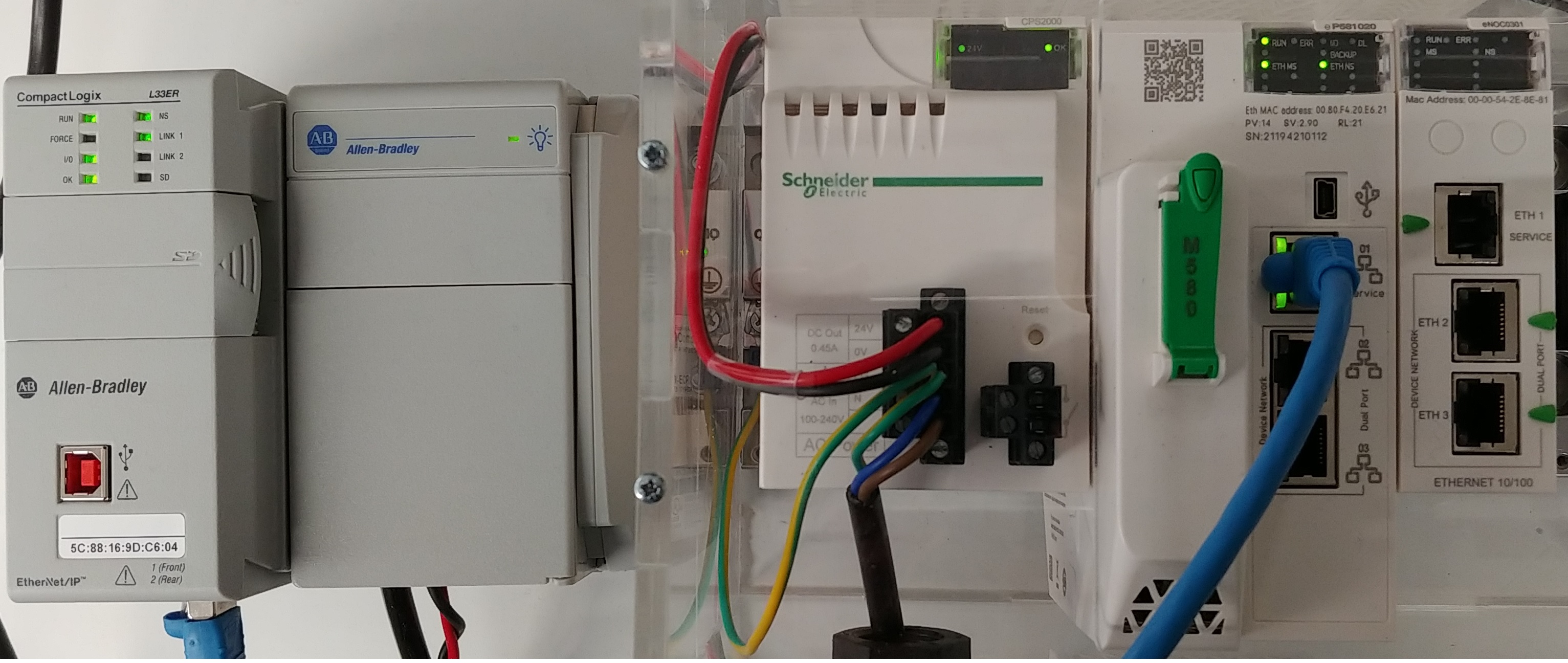}}
\caption{PLCs: CompactLogix 5370-L3 and Modicon M580}
\label{fig5}
\end{figure}

In order to achieve FDIA on our digital twin system, multiple well-known security analysis tools were used:
\begin{itemize}
\item 
Kali Linux 2020.2: Kali Linux was chosen as our attacker machine environment because various popular penetration testing tools are integrated into it.
%Kali Linux is an operation system which is integrated with many hacking and penetration tools. It is used as the main attacking environment. TJ
\item 
Ettercap 0.8.3: Ettercap is a well-known MitM attack tool and it has been used in multiple papers discussed in Section \ref{section6}. Etterfilter, a filter compiler for Ettercap content, is used to manipulate the network packet contents in real-time. 
%Ettercap is used for launching the man-in-the-middle (MITM) attack. Ettercap filter function is used to manipulate the network packet contents in real-time. TJ
\item 
Wireshark 3.2.4: A widely-used tool for capturing and analyzing network packets.
%Wireshark 3.2.4: Wireshark is used to capture and dissect packets transfer between HMI and PLCs.
\item 
Kdiff3 0.9.98: Kdiff3 is an open-source software that can be used to compare extracted packet contents.
\end{itemize}

\subsection{Verification: Application of the Analytics Framework}
We discuss the application of each step of our framework onto the digital twin systems:
\begin{description}
\item 
\textit{Getting into OT network:} Since this step is out of our scope, we directly connected our attack machine to a local network switch.
%Step 1: Get into the OT network by connecting the attack machine to local network switch.
\item 
\textit{ARP poisoning:} After launching Ettercap in attack machine, we scanned the OT network and targeted HMI and PLCs for the ARP-Poisoning attack. Then, a MitM attack was started against the experimental system.
%Step 2: Launch the Ettercap in Kali linux, scan the network for hosts, add HMI and PLCs IP addresses as the targets and start ARP Poisoning.
\item 
\textit{Packet sniffing:} We captured all network packets coming into the attack machine's network card using Wireshark.
    %Step 3: Start capturing and monitoring the communication packets by using Wireshark.
\item 
\textit{Packet classification:} The captured packets were sorted according to the packets' protocols: CIP for CompactLogix 5370-L3 and Modbus for Modicon M580. In the case of other protocol packets, we ignored them as they did not contain meaningful information for our scenario.
% TJ Camera 'do' -> 'did'
%Step 4: Sort the captured packets according to the protocols: CIP and Modbus. TCP packets and packets from other protocols are ignored as they do not contain useful information in our case.

\item
\textit{Sorting periodical sending order:} We found that the HMI and PLCs send sequences of request and response packets that have the same packet length individually by CIP and Modbus after inspecting captured network packets for the same protocol packets. After that, we sorted the sequences by protocols.

\item 
\textit{Grouping by the same packet length:} We grouped the packet lengths of request and response packets individually by CIP and Modbus. After that, packets with the same length in the same protocol were grouped.
%Step 5: Conduct the analysis of CIP and Modbus packets individually. For the same protocol packets, find out the request and response packets send sequence. Then sort them according to the packet lengths.

\item 
\textit{Extracting data field:} After opening the captured packets with Wireshark, we copied the `Modbus$\rightarrow$Data' field and `CIP Class Generic$\rightarrow$Command Specific Data$\rightarrow$Data' field for each protocol into empty files. The created files maintained by the groups were sorted by the packet length at the previous stage.
%Step 6: Copy the hex format payload data of the packets into text files, each file contains different packet data under the same packet length, then use Kdiff3 to visually locate the data differences.

\item
\textit{Finding changed field:} By comparing the new files created from the `Extraction of data field' step using Kdiff3, we found the changing data field. The compared files were chosen from the same group. 

\item 
\textit{Finding the range of each variable and sequential relationship:} By comparing each changeable field with the same field of other packets within the same group, a range of changeable bytes or field can be figured out. In the case of time sequence for changeable fields, we figured out it by comparing the timestamps of the packets from other groups.
%Step 7: From step 6, the changing fields can be obtained, then conduct further analysis for finding the changing rang of the targeted bits and sequential relationship between these bits.

\item
\textit{False data injection attack:} After setting up Etterfilter (one of the functions within Ettercap) with targeted IP and port information of HMI or PLCs for intercepting required  packets, we were able to manipulate the targeted changeable data field. To overcome the functional limitation of Etterfilter, which can only replace an exact value with a substitute, we developed a Python program that can be called by the Etterfilter. The program keeps changing the targeted data field with prepared fake payloads within the changeable range figured out from previous step. With the combination of Etterfilter and the Python program, the FDIA appears the same as the real data changes from the perspective of the victim PLCs.
%Step 8: Prepare the Ettercap filter with the corresponding IP and port information for intercepting the required packets.
%\item Step 9: Prepare a modified fake payload in a python program and write the program so that it keeps changing the specific data fields in order to imitate how the real data changes.
%\item Step 10: Dynamically replace the real-time packet payload with the prepared one by using a combination of the Ettercap filter and the Python program.
\end{description}

\subsection{Findings}
%TJ
%If we have the request packet with length 100 and two corresponding response packets with length 200 and 150, the packet with length 100 is sent first by HMI. Then it receives two responses with length 200 and 150 respectively. The packet sequence pattern is: 100$\rightarrow$200$\rightarrow$150$\rightarrow$100$\rightarrow$200... The red color numbers shown in the table have the changeable property, and the rest of the bits are fixed all the time. After analyzing, these varied bits are found to be used to control different components of our simulation system (e.g. wind speed, barrel position, indicator light, etc.). By modifying these changeable bits using Ettercap, the changing behaviours of these bits can be discovered which are used to deceive HMI with fault data.\\ \\
%The varied payload bits after analyzing are shown in Table X:
Through the process up to the step `Finding changed field' on the captured network data between the HMI and the PLCs, we identified changeable data fields described in Table~\ref{tab1}. The numbers in red in the table indicate the changeable fields, while the rest of the numbers are fixed all the time. 

%\begin{table}[h!]
\begin{table}[htbp]
  \caption{Changeable Data Field}
  \begin{center}
    \begin{tabular}{c|c}
      \textbf{Origin} & \textbf{Correlated payload bits}\\
      \hline
      Rockwell & \textcolor{red}{00}020000...05001d...0600880b...0000080000 \\
      Rockwell & 000200\textcolor{red}{00}...05001d...0600880b...0000080000 \\
      Rockwell & 00020000...0500\textcolor{red}{1d}...0600880b...0000080000 \\
      Rockwell & 00020000...05001d...0600\textcolor{red}{880b}...0000080000 \\
      Rockwell & 00020000...05001d...0600880b...\textcolor{red}{00}00080000 \\
      Rockwell & 00020000...05001d...0600880b...00000800\textcolor{red}{00} \\
     Schneider & \textcolor{red}{01}0044000000f77f41440e06 \\  
     Schneider & 01\textcolor{red}{00}44000000f77f41440e06 \\
     Schneider & 0100\textcolor{red}{44}000000f77f41440e06 \\
     Schneider & 010044000000f77f\textcolor{red}{4144}0e06 \\
     Schneider & 010044000000f77f4144\textcolor{red}{0e06} \\
     HMI & 01010004004e03206b25000e00010001\textcolor{red}{ff} \\
     HMI & 01010004004e03206b25000e00010001\textcolor{red}{fe} \\
    \end{tabular}
  \label{tab1}  
  \end{center}
\end{table}

\section{Result Analysis and Countermeasures} \label{section5}

Table~\ref{tab2} shows the matching data fields for each HMI server item. Using the same way in which an adversary could monitor an exploited HMI server, we figured out these data fields by monitoring HMI screen with data in Table~\ref{tab1}. As highlighted in Section~\ref{Monitoring_HMI}, the high rate of RAT usage in the field for ICS implies high probability of a successful attack using our methods.

\subsection{Attack Result}
Based on the findings described in Table~\ref{tab1}, we proceeded with the two remaining steps in the framework. As a result, we successfully manipulated the values in the highlighted fields using Etterfilter. Table~\ref{tab2} shows the original and rigged value of each component.

\begin{table}[h!]
  \caption{Attack Result - Original and Rigged Values For Each Component}
  \begin{center}
    \begin{tabular}{c|c|c}
      \textbf{HMI components} & \textbf{Original value} & \textbf{Modified Value}\\
      \hline
     System switch & \textcolor{red}{00}020000 & \textcolor{red}{01}020000 \\  
     Conveyor belt speed & 0500\textcolor{red}{00}00 & 0500\textcolor{red}{1d}00\\
     Panel turbine speed & 0600\textcolor{red}{0000} & 0600\textcolor{red}{ffff} \\
     Place a new barrel & 0200\textcolor{red}{00}0003 & 0200\textcolor{red}{01}0003 \\
     Brake light & \textcolor{red}{01}0044...f77f41440e06 & \textcolor{red}{00}0044...f77f41440e06 \\  
     Contractor light & 01\textcolor{red}{00}44...f77f41440e06 & 01\textcolor{red}{01}44...f77f41440e06\\
     Wind speed & 0100\textcolor{red}{44}...f77f41440e06 & 0100\textcolor{red}{50}...f77f41440e06\\
     Turbine speed & 010044...f77f4144\textcolor{red}{b80b} & 010044...f77f4144\textcolor{red}{0000}\\
    \end{tabular}
    \label{tab2}
  \end{center}
\end{table}

The sequence changing of the data fields depicts two cases (see Figure \ref{fig4}):
\begin{description}
\item 
\textbf{Case 1: System switch is turned `ON'.}\\
\textit{`Wind speed/Turbine speed/Panel turbine speed'$\rightarrow$`Brake light'$\rightarrow$`Contractor light'$\rightarrow$`Place a new barrel'}\\
%RK: Urgent: Please add a paragraph to explain what these switching on and off mean! 

%When the system switch is turned on, all the turbine speed displays started to gradually increase until the value reaches 3000 RPM. After that, the brake light will be changed from `on' state to `off' state and the contractor light will be changed from `off' state to `on' state.
%Case1: The turbine speed is 0 RPM and system switch is off.\\
%When the system switch is turned on, all the turbine speed displays start to gradually increase until the value reaches 3000 RPM. After that, the brake light will be changed from on state to off state and the contractor light will be changed from off state to on state.

%\item 
%Case2: The turbine speed starts increasing but have not reached the maximum value yet, and the system switch is on.\\
%When the system switch button is turned off, all the turbine speed displays are changed to 0 RPM at once. The light indicators remain the same.

\item
\textbf{Case 2: System switch is turned `OFF'.}\\
`Wind speed/Turbine speed/Panel turbine speed'$\rightarrow$`Brake light'$\rightarrow$`Contractor light'
%Case3: The turbine speed reaches maximum value, and system switch is on.\\
%When the system switch button is turned off, all the turbine speed displays start to gradually decrease until the value reaches 0 RPM. After that the brake light will be changed from off state to on state and the contractor light will be changed from on state to off state.

%RK: Urgent: Please add a paragraph to explain what these switching on and off mean! 

\end{description}
The `system switch' is for turning `on/off' our ICS HMI system. As shown in Figure \ref{fig4}, the electricity for our conveyor belt simulator (i.e. top half of Figure \ref{fig4}) is generated from our wind-farm simulator (i.e. bottom half of Figure \ref{fig4}), the turning `On' of the `system switch' means that the wind turbine is switched on and ready to generate electricity in the wind-farm simulator. On the contrary, 'system switch' `Off' means stopping the wind (and hence the electricity generation) for the wind-farm.

The data field change ranges are as follows:
\begin{itemize}[itemsep=1mm, parsep=0pt]

\item System switch: `0x00' - `0x01'.
\item Panel turbine speed: `0x0000' - `0xb80b'.
\item Turbine speed: `0xff39' - `0x3b45'.
\item Brake light: `0x00' - `0x01'.
\item Contractor light: `0x00' - `0x01'.

%RK: Urgent: Please add a paragraph to explain what these numbers mean!
\end{itemize}
From our framework, we managed to figure out 
the lower/upper bound numbers of the range
% TJ Camera
and scale for each component from gathered network packets -- simulating successful heuristic inference. 
The lower bound number for each data field is the smallest value of the data field we have monitored from response packets for the HMI, 
the upper bound number is the biggest value, and scale stands for the incremental step. 
% TJ Camera add 'and scale' and 'and scale numbers'
%and scale we have monitored are the upper bound and scale numbers for each changeable data field.
Once an attacker knows these bounds, the attacker can stealthily launch a range of deception attacks to the HMI and the PLCs. For example, an attacker could alter the appearance of the HMI displayed value of the turbine speed even though the actual PLC linked to the wind turbine is not turning. Another example is when the attacker can trick the conveyor belt into moving faster than normal, thereby disrupting the manufacturing process and causing the barrel to spill over from the conveyor belt.

\subsection{Time Complexity of Packet Analysis}
To prove the reproducibility of the proposed heuristic inference attack procedure, we analysed the task complexity of the proposed framework. High complexity tends to require longer time and expertise for performing analyses -- making the procedure difficult to reproduce. On the other hand, lower complexity procedures are easier to reproduce. 

We refer to \cite{LIU2012553} for the calculation for task complexity. \cite{LIU2012553} used `size' as a complexity dimension of the proposed framework because the packet analysis task components are sorting and grouping network packets, and finding data fields from the same packets. Due to the same reason, the sole complexity contributory factor in \cite{LIU2012553} is `Input Quantity'. In the same manner, the task complexity of the heuristic inference process can be presented using the time complexity of the algorithm composed of the inference steps.

% TJ Camera add 'Input' and 'Output'
\begin{algorithm}
\SetAlgoLined
 \textbf{Input:} Captured network packets between targeted HMI and PLCs
 
 \textbf{Output:} Identification of changeable data fields, low/upper bound numbers and scale for each field, and sequential relationship
 
 Classification of IP Addresses assigned to each PLC\;
 \While{for each IP Address}{
  Inference Step 1: Sorting periodical sending order\;
  Inference Step 2: Grouping by the same packet length\;
  Inference Step 3: Extracting data field\;
  \While{for each group of same packet length}{
    Inference Step 4: Finding changed field\;
   \While{for each particular changeable data field having a difference}{
     Inference Step 5: Finding the range of each variable and sequential relationship\;
   }
  }
 }
% \KwResult{Write here the result }
 \caption{Procedure of Packet Analysis proposed}
 \label{alg:pro}
\end{algorithm}

% TJ Camera
The procedure of our packet analysis can be represented as Algorithm~\ref{alg:pro} using pseudocode. From the algorithm, we can see that packet analysis is a recursive process and thus its time complexity depends on the three looping parameters: (1) the number of PLCs (i.e., the number of IP Addresses), (2) the number of different packet length, and (3) the number of changeable data fields. These parameters are determined by the specific implementation of each ICS. In practice, the overall time consumption is not high, since each action takes short time due to existing tool support. In particular, the repetition number of `Grouping by the same packet length' and `Comparison of extracted data field' depend on the number of I/O modules per each PLC and the variation range of I/O modules, hence the number for reiteration in practice is typically low. Therefore, our proposed framework will be feasible.

\subsection{Validation: Dataset from different environmentt} \label{SUTD}
% TJ Camera
To validate our analysis framework and prove the feasibility of our approach, we applied our approach to analyse an entirely different dataset from a water cyber-physical system setup from the Singapore University of Technology and Design (SUTD). After analysing the SUTD captured network packets, we found that the SUTD system contains sixteen PLCs and one HMI, which is larger than ours in its scale.. Each PLC has fixed lengths of HMI request and PLC response messages that have been used for heuristic inference. 
% TJ Camera
After completing all packet content analyses, we observed that their PLCs have similar changing behaviors with ours. By applying the same packet analysis techniques on the data fields, we identified some data fields which may relate to system settings and value modifications. However, without checking the HMI, the purposes of the data fields cannot be validated. Hence it requires a long time to accumulate and analyse data in order to increase the accuracy of the predicted results. Parts of the analysed results are listed below in Table~\ref{tab3}. Our experiment on this dataset demonstrates  that key components of our approach (e.g., packet classification and inference) are scalable when applied in complex ICSs. 

\begin{table}[h!]
  \caption{heuristic inference result on SUTD data}
  \begin{center}
    \begin{tabular}{c|c|c}
      \textbf{Packet length} & \textbf{Data payload} & \textbf{Predicted purpose}\\
      \hline
     183 & 870508xxxxxxxx & System parameters change \\  
     520 & 10083010\textcolor{red}{0}84030 & Function On \& Off\\
     130 & 3350524f54 or 334354524c & System status toggle \\
    \end{tabular}
 % \end{center}
 
%\end{table}
\vspace{10pt}
%\begin{table}[h!]
 % \begin{center}
    \begin{tabular}{c|c}
      \textbf{Packet length} & \textbf{Correlated payload bits}\\
      \hline
     183 & 870508\textcolor{red}{4374ece7} \\  
     520 & 010\textcolor{red}{0}xxxxc5b26d3b007c \\
     520 & 0100xxxxc5\textcolor{red}{b26d3b007c} \\
     130 & 4544\textcolor{red}{3350524f54} \\
     130 & 4544\textcolor{red}{334354524c} \\
    \end{tabular}
     \label{tab3}
  \end{center}
\end{table}
 
\subsection{Countermeasures}
The customary countermeasure for our suggested attack scenario is to implement secure communication channels and continuous monitoring of the OT network. Cryptography can be used to ensure confidentiality of communication between HMI and PLCs, 
% TJ Camera 'will not able to' -> 'will not be able to'
so that adversaries will not be able to analyse internal packets (as compared to plaintext in most current ICS setups). By monitoring OT network, the system administrators will be able to block network connection requests except for trusted devices in the internal network and reduce the risk surface. Hence, the devices connected to the internal network should be under surveillance for anomalous/deviant messages.

Another countermeasure to obstruct attackers is to Increase the time complexity by obfuscating program source code. As shown in `Algorithm~\ref{alg:pro}', introducing an addendum on the loop conditions is an effective way to complicate the analysis. Adding some fake PLCs using soft PLC in the OT network is the another effective way to increase the outer most loop. If a communication channel is divided into ports, there will be an additional loop under each IP address.

To raise the complexity for the packet analysis task further, we could also introduce a pseudorandom number generator into the ICS implementation. If an engineer inserts an additional meaningless field using a pseudorandom number generator, or further bytes which are coded to a changeable data field with the random number, the time complexity of packet analysis will definitely increase (according to Algorithm~\ref{alg:pro}).

\section{Related Work} \label{section6}
A number of works focused on protecting ICS by detecting attack attempts on the infrastructure but did not provide sufficient information for reproducibility. For example, the authors of ~\cite{10.1145/2664243.2664277,gao2014cyber,he2019power} demonstrated test results of their suggested protection methods, but failed to discuss any information on (re)producing the used attack data. Lan et al. \cite{9045827} introduced their tools used for MitM attacks, and \cite{li2018two} briefly described an attack scenario using tools such as Ettercap, Pypacker, and Wireshark.

Amin et al. \cite{10.1145/1755952.1755976} focused on generating deceptive data and tested the data using a developed SIC software to a local canal in Southern France.  

 Direct PLC attack scenarios have also been considered but most of them required sizeable time and domain expertise before attacks could be launched successfully –– a key difference with our proposed approach which requires no prior domain knowledge in automation. For example, Wallace and Atkison demonstrated a scanning attack, an attack extracting tags, and a CPU toggling attack for a model of Siemens S7 line inside a testbed OT network \cite{wallace2013observing}. The work \cite{klick2015internet} demonstrated a tactic of how adversaries were able to access the PLCs positioned in OT network. After finding a PLC which is connected to the Internet, a SOCKS proxy was injected to reach internal PLCs and committed a download attack to the Siemens PLC with pre-prepared PLC code.

Demonstrating download attacks on PLCs, Yoo and Ahmed performed experiments controlling logic injection attacks by analysing packets on Schneider Electric and Allen-Bradley PLCs \cite{yoo2019control}. Kalle et al. \cite{kalle2019clik} succeeded in downloading new firmware, which could be modified by researchers using a self-implemented decompiler on a Schneider Electric's PLC via direct access to the device.

As a type of Denial-of-Service (DoS) attack, \cite{wang2018access} showed an attack by inverting values from read/write registers using Python code implemented by researchers. Another group of researchers suggested a markup language dedicated to ICS attacks \cite{kleinmann2017stealthy}. With their proposed language, Kleinmann et al. conducted MitM attack including zero paddings and shifting register addresses for network data packets between HMI and PLCs.

Yrlmaz et al. \cite{yrlmaz2019cyber} conducted a DoS attack on Siemens and Schneider Electric PLCs. They also conducted a MitM attack which achieved gathering network packet data using a combination of tools such as Ettercap and Wireshark. Kang et al. also used Ettercap and Wireshark for data injection attacks in \cite{kang2015investigating}.

The CompactLogix 5370 used for our simulation system had also been exploited by Pavesi et al. in \cite{pavesi2019validation}. These researchers developed a tool for a SYN injection attack for the TCP protocol and they conducted the attack in the OT network of their lab environment.

Miciolino et al. \cite{7377479} used Ettercap and Etter-Filter to conduct a FDIA exploit on a water system testbed using Schneider Electric's Modicon M340. Using their expertise, researchers were able to modify command and send false responses. The  researchers also conducted ping flood and Modbus flooding attack with FDIA in \cite{bernieri2017monitoring}. They evaluated a Fault Diagnosis module and Intrusion Detection System in the same testbed environment.

\section{Concluding Remarks and Future Work} \label{section7}
%the experiments are not concluded yet so it is more appropriate to say 'concluding remarks'.
We proposed an analytics framework using a heuristic inference approach to produce False Data Injection Attack (FDIA) data in an ICS network. The proposed framework targets the communication channel between the supervisory control network and the control network typically found in many ICS scenarios. Our approach leverages a well-known network penetration tool and network analysis tool to commit a man-in-the-middle (MitM) attack and ICS network packet analysis. 

The major advantage of our framework is that no prior expertise for ICS and automation engineering is required to follow the suggested steps in our framework. We demonstrated an attack scenario on a system simulating a wind-farm powering a barrel conveyor belt in a manufacturing plant. The experiment showed that our framework was able to make stealthy deception attacks achieved by a non-ICS expert. The application of our framework onto a second dataset validated our approach and showed the versatility of our analytics framework. 

For future work, we will continue to conduct further research on the countermeasures proposed earlier in this paper. Based on the data produced by our framework, future research on preventative measures would be feasible and aligns well to the ultimate goal of our framework. As an extension of our approach, we will conduct further experiments of our framework on an ICS with a fake PLC or add meaningless data fields to experiment control data integrity and resiliency.

\section{Acknowledgements} \label{section8}

We thank the anonymous reviewers for their valuable feedback. We also acknowledge Bob Stokes, Alex Ladur, and Jean-Paul Mondon-Ballantyne from Combined Technologies (CTEK) Ltd for providing their domain expertise during the setup of the ICS systems and scenarios at the UQ Energy TestLab (Website:  https://energy-testlab.lab.uq.edu.au). CTEK also coordinated our disclosures of vulnerabilities discovered by our framework. The data used in Section~\ref{SUTD} was provided by the ``iTrust, Centre for Research in Cyber Security'' at the Singapore University of Technology and Design. We also thank the UQ Cyber Security strategic funding for the procurement of the ICS systems.

% TJ start
\bibliographystyle{IEEEtran}
\bibliography{Paper_IEEE_2020_1}
% TJ end
\end{document}